\RequirePackage{lineno} 
\documentclass[final,authoryear,12p]{elsarticle}

\usepackage{epsfig}
\usepackage{verbatim}
\usepackage{amssymb}
\usepackage[%
a4paper=true,%
breaklinks=true,%
colorlinks=true,%
pdfauthor={First Author et al.},%
pdftitle={Template for manuscripts in Advances in Space Research}%
]{hyperref}

\journal{Advances in Space Research}

\begin{document}
\begin{frontmatter}



\title{Measurements of the cosmic ray spectrum and average mass with IceCube}


\author{ Shahid Hussain\corref{cor} }
\address{Department of Physics and Astronomy, University of Delaware}
\cortext[cor]{Corresponding author}
\ead{shahid@bartol.udel.edu}
\author{for the IceCube collaboration}
%
 \ead[url]{http://www.icecube.wisc.edu}

\begin{abstract}
Located at the South Pole, IceCube is a particle-astrophysics observatory composed of a square-kilometer surface air shower array (IceTop) and a 1.4 km deep cubic-kilometer optical Cherenkov detector array. We review results of measurements of the cosmic ray spectrum and average mass in the energy range 1 PeV to 1 EeV.

\end{abstract}


\end{frontmatter}

\parindent=0.5 cm

\newpage
\section{Introduction}
\subsection{Cosmic rays }

Cosmic rays are naturally produced in the universe, having a spectacular energy range that runs  from very low energies to extremely high energies ($\sim 10^{21}$\,eV) that are beyond the reach of man-made particle accelerators. Due to their natural origin and wide energy range, they are crucial to enhance our understanding of  the universe. Although it has been a century since the discovery of cosmic rays, there are several unanswered questions that make cosmic ray physics very interesting and an active field of research. These questions are related to cosmic ray identity, origin, acceleration mechanism, and scaling of fundamental interactions with energy. For recent reviews of cosmic rays, see \citep{CRRevKW, CRRevGS}.

Regarding the sources of cosmic rays in our galaxy, the energy dynamics of Supernova Remnants (SNRs) makes them a strong candidate that could produce cosmic rays reaching the Earth up to energies around $10^{15}$-$10^{18}$ eV; to accelerate cosmic rays to even higher energies,  one needs extreme environment that might not be possible to achieve in our galaxy; however, exotic extragalactic objects like Active Galactic Nuclei (AGN) and Gamma Ray Bursts (GRB) are possible candidates for the highest energy cosmic rays. 

Balloon and satellite measurements of cosmic rays have been performed to maximum energies of the order of $\sim100$ TeV  \citep{CRRevM}. At higher energies, the cosmic ray fluxes are too small for direct detection and large surface detector arrays are used to detect cosmic ray air showers. As compared to direct detection, the complexity of air showers makes indirect detection and composition analysis a lot more difficult. Therefore, several  independent experiments and analysis methods have been invented to improve our knowledge of cosmic ray composition and spectral details above energies where direct measurements run out of statistics. 

The cosmic ray spectrum follows a power law, implying they are not black body radiation. However, the power law has breaks which could be attributed to  appearance and disappearance of different types of cosmic rays sources. Detailed study of the spectral structure above a PeV is one of the major goals of cosmic ray surface detectors. 

\subsection{Cosmic ray air showers in IceCube}
\begin{figure}
\begin{center}
\includegraphics*[width=16cm,angle=0]{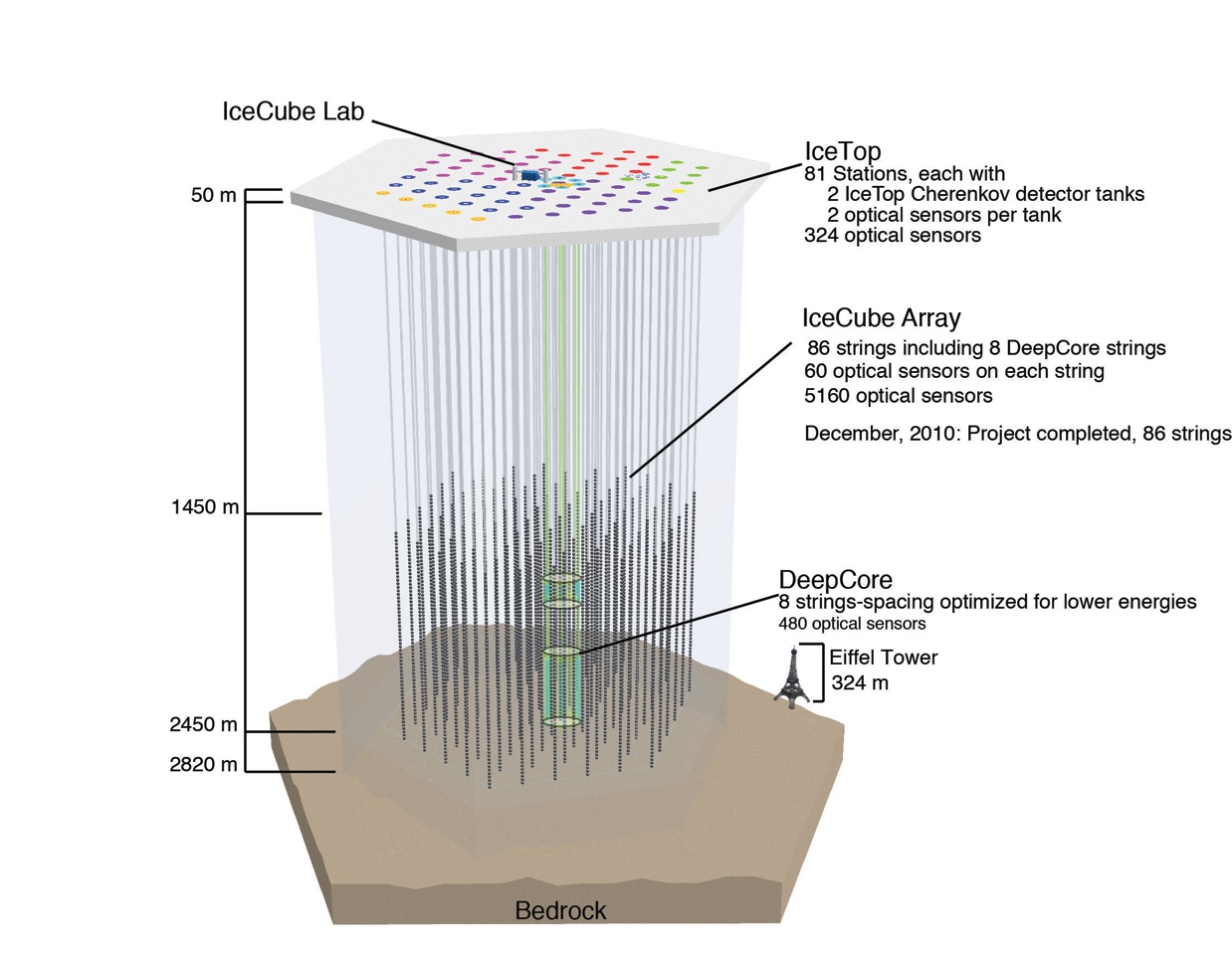}
\end{center}
\caption{Drawing of the IceCube observatory (side view).}
\label{fig:icecube}
\end{figure}

 IceCube is located at the South Pole. As shown in figure~\ref{fig:icecube}, IceCube is a cosmic ray and neutrino observatory with IceTop surface area of a square-kilometer and in-ice instrumented volume of a cubic-kilometer. For technical details, see \cite{icecubeDAQ}, \cite{icecubePMT}, and \cite{icetop}. While the ice-filled tanks of the surface array detect Cherenkov light from muons and electromagnetic particles, the 1.4 km deep in-ice array detects Cherenkov light produced by only the high energy muons (around 300 GeV and above) present in the shower core. This feature makes IceCube a unique cosmic ray detector as the combined information from surface and in-ice arrays can be useful in determining detailed structure of cosmic ray spectrum and composition. Below, we briefly describe the ability of IceCube to detect air showers.

IceTop \citep{icetop} has an elevation of 2835 m above sea level that corresponds to an average atmospheric depth around 680 $\mathrm{g/cm^3}$. Cosmic rays produce air showers when they enter our atmosphere; as an air shower develops, the total number of particles in the air showers reaches a maximum and then decreases again. For a given primary cosmic ray type the depth of shower maximum depends on the energy of the primary cosmic ray; from a few PeV to EeV energies, the shower max falls  roughly within 20\% above the depth of 680 $\mathrm{g/cm^3}$. This makes IceTop especially useful to obtain a better energy resolution of the primary cosmic rays.

At the surface, the Cherenkov light is produced by muons and electromagnetic particles of the air shower as they pass through the ice-filled IceTop tanks. IceTop tanks have cylindrical shape, filled with 90 cm of transparent ice. The tank walls are lined with reflective material and tank tops are level with surface (i.e. the tanks are buried under ice). There are two DOMs (Digital Optical Modules) in each tank. Each DOM has a trigger logic circuit, ATWDs (Analog Transient Waveform Digitizers) with 3.3 ns bins, and a 10 inch PMT (Photo Multiplier Tube) to collect Cherenkov light produced by charged particles  in the tank. The two DOMs of a tank are set to two different gains to have a broader dynamic range; one DOM is set to High Gain (HG) and the other to Low Gain (LG).

\begin{figure}

\begin{center}
\includegraphics*[height=10cm,width=15cm,angle=0]{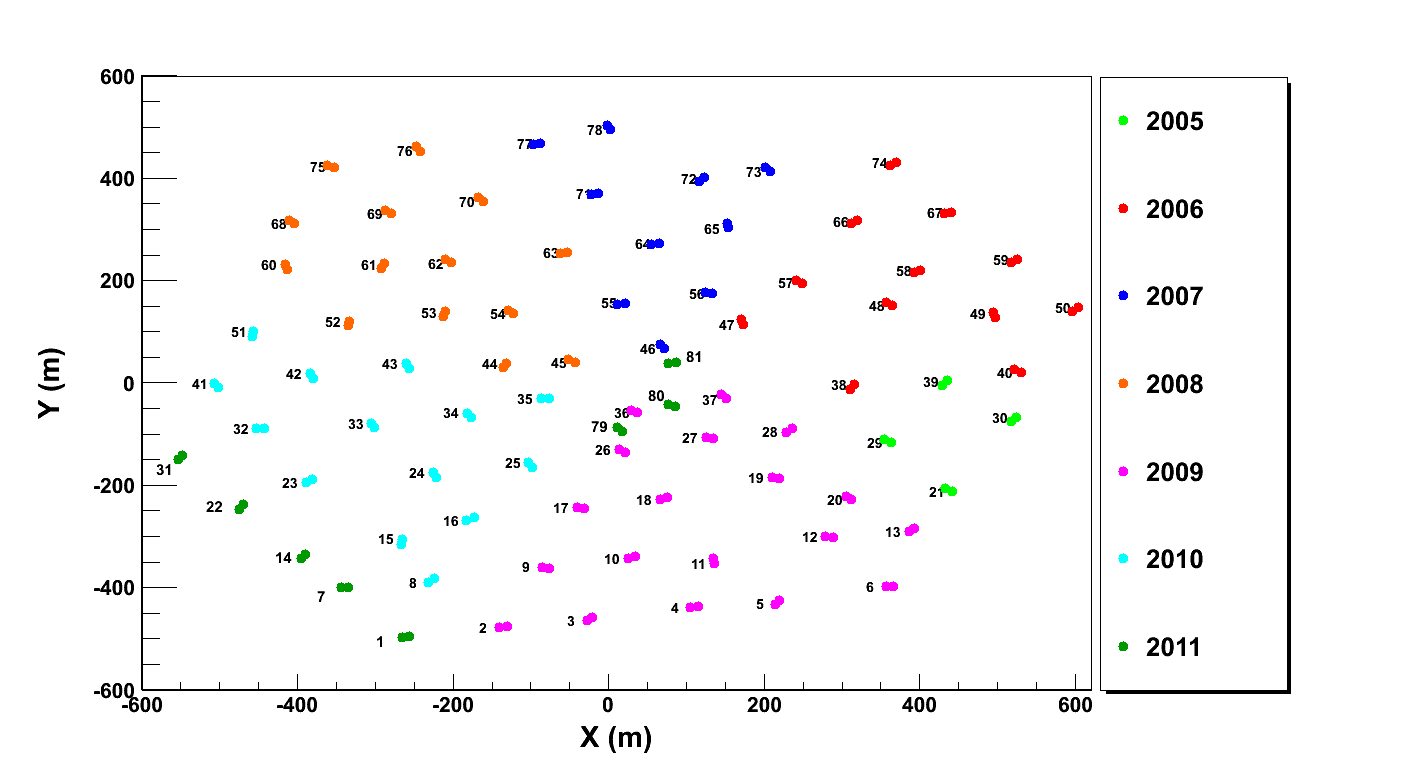}
\end{center}
\caption{IceTop geometry: Each pair of colored dots represents two IceTop tanks of a station and the colors correspond to deployment years.}
\label{fig:icetop}
\end{figure}

As shown in figure~\ref{fig:icetop}, the IceTop tanks are in pairs, separated by 10 m; each pair forms a station. The main array inter-station spacing is 125 m and there are three infill stations (labeled 79, 80, and 81 in ~\ref{fig:icetop}) that reduce the spacing between some of the tanks near the center of the array. The infill stations and their neighboring stations bring the IceTop threshold to cosmic rays of energies down to a few hundred TeV. The lower threshold makes it possible to compare IceTop cosmic ray measurements with the direct measurements from balloon and satellite experiments.

The tank signals are calibrated and have a time stamp  and integrated charge. The calibrated tank signals have units of VEM (Vertical Equivalent Muon); 1 VEM is the signal produced by an energetic vertical muon passing through an IceTop tank. The process to achieve these calibrated signals is explained in detail in  \citep{icetop}. Given the measured signal and trigger time information from the IceTop tanks, IceTop software reconstructs air shower core location, direction, and lateral distribution of signals in the tanks. Simulations are used to estimate the energy of primary cosmic rays by relating the expected tank signal at a lateral distance of 125 m ($S125$) from the shower core to the true primary cosmic ray energy. The dependence of this relation on the type of primary cosmic ray depends on the energy and zenith angle of primary cosmic ray; parameterizations of the primary energy versus $S125$ are obtained from simulations of different cosmic ray types.

\begin{figure}
\begin{center}
\includegraphics*[width=10cm,angle=0]{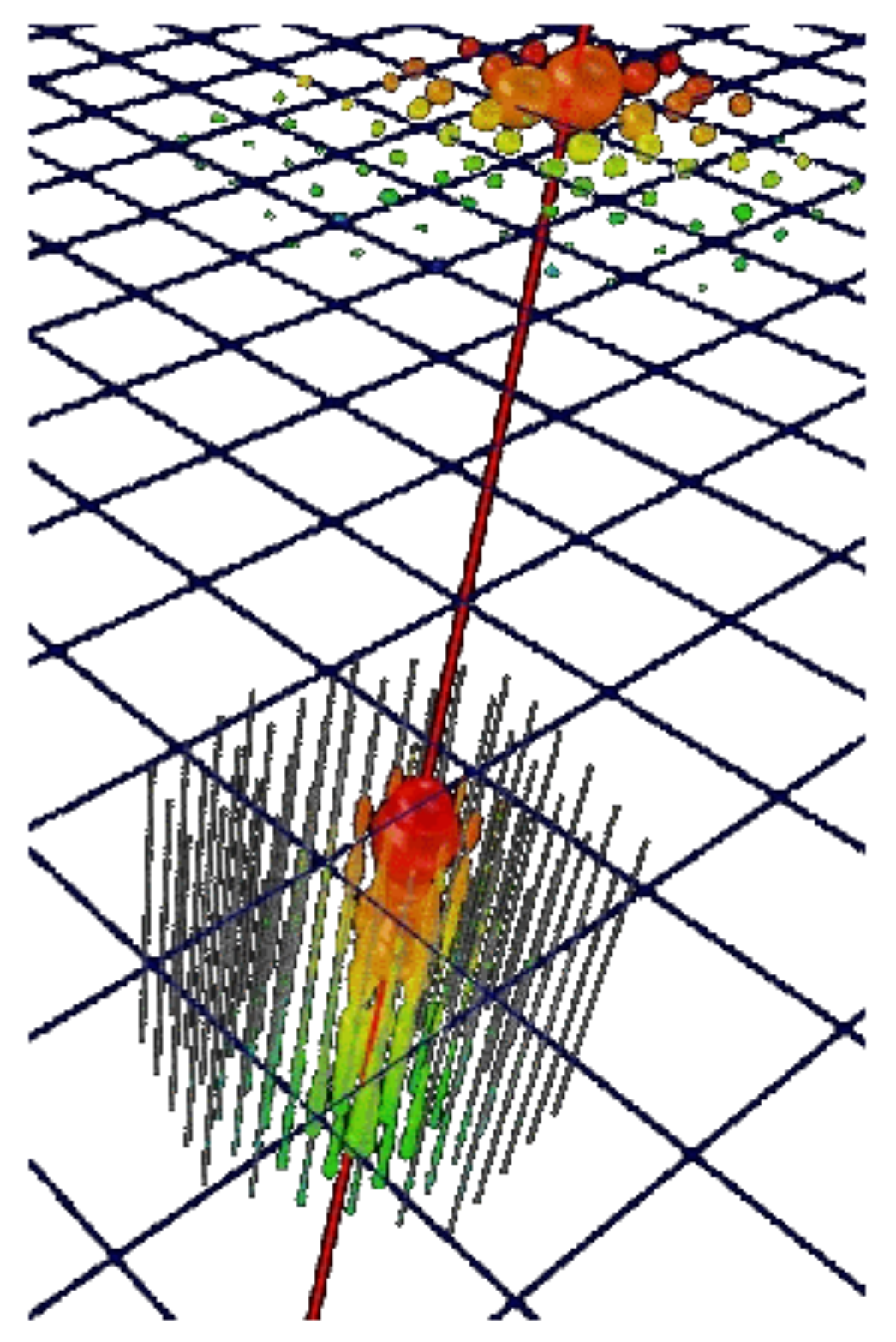}
\end{center}
\caption{An example of a real air shower event in IceCube that triggers both surface and in-ice arrays. Size of the colored blobs is proportional to logarithmic signal pulse-charge.}
\label{fig:eventIC}
\end{figure}

\begin{figure}
\begin{center}
\includegraphics*[height=10cm,width=15cm,angle=0]{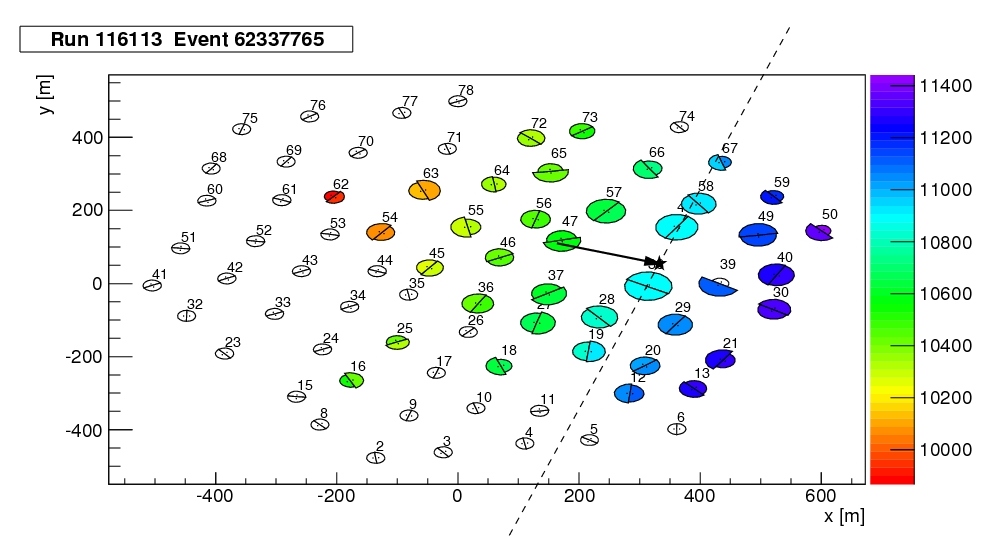}
\end{center}
\caption{An example of a real IceTop air shower event. Different colors give the tank pulse trigger time and the size of these colored blobs is proportional to logarithmic signal pulse-charge. The black arrow in the figure panel gives reconstructed shower direction, and the black star at the end of arrow shows reconstructed core position.}
\label{fig:eventIT}
\end{figure}

\begin{figure}
\begin{center}
\includegraphics*[width=15cm,angle=0]{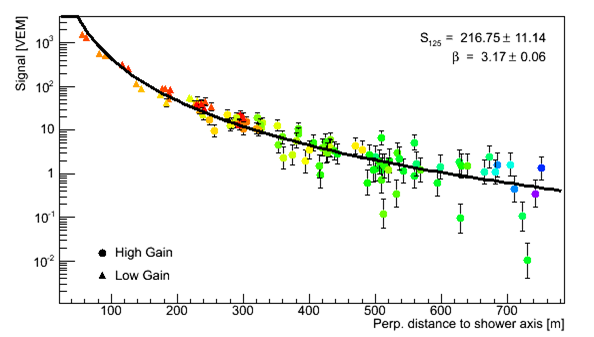}
\end{center}
\caption{Tank signal vs reconstructed lateral distance for a real air shower event in IceTop: The colors correspond to DOM trigger times and red is the earliest. The triangle shape represents Low Gain and the circles represent High Gain DOMs. The black curve is reconstructed fit result. Also shown in the figure panel are shower size $S125$ and slope $\beta$  parameter values of the fit. The reconstruction algorithm is explained in \cite{icetop}.}
\label{fig:qfit}
\end{figure}

Figure~\ref{fig:eventIC} shows the side view of an actual cosmic ray air shower event in IceCube and figure~\ref{fig:eventIT} shows the top view of an actual event in IceTop. Figure~\ref{fig:qfit} shows the tank signals vs the reconstructed lateral distance between the tanks and reconstructed shower core; near the core, only Low Gain DOMs are present and as we move away from the core, only High Gain DOMs are present at large enough distances; this is because High Gain DOMs are saturated at smaller distances and cannot be used for reconstruction, while at larger distances the signals are small and cannot trigger Low Gain DOMs.  

\section{Cosmic ray spectrum and average mass measurements with IceCube}

\begin{figure}
\begin{center}
\includegraphics*[width=15cm,angle=0]{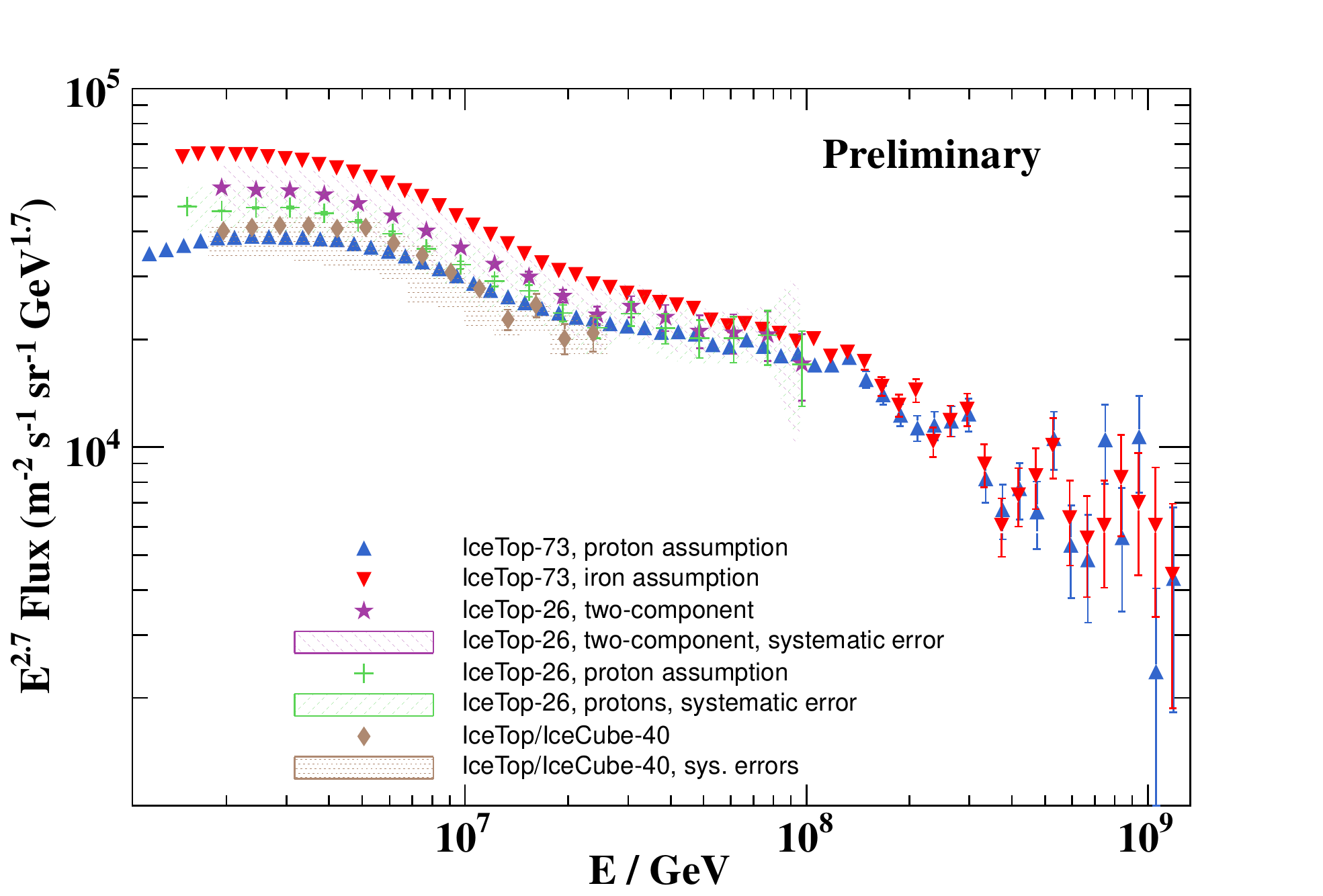}
\end{center}
\caption{Comparison of spectra resulting from different analyses in IceCube: IceTop-only analysis using 26 stations (IceTop-26), IceTop-only analysis using 73 stations (IceTop-73), and IceCube (in-ice and IceTop combined) analysis using 40 IceCube strings (IceCube-40).}
\label{fig:specIC}
\end{figure}

Figure~\ref{fig:specIC} gives a comparison of the all particle cosmic ray spectra obtained by different IceCube analyses: IceTop-26 \citep{IT26}, IceTop-73 \citep{IT73}, and IceTop/IceCube-40 \citep{IC40}. Below we briefly summarize the three analyses.

The IceTop-26 spectrum analysis results shown in figure~\ref{fig:specIC} are based on data obtained by 26 IceTop stations \citep{IT26}; in-ice array was not utilized for the analysis. The analysis gives the all-particle spectrum in the energy range between 1-100 PeV from data taken between June and October 2007. The major source of systematic error for this analysis is the unknown composition of the primary cosmic rays. The analysis was performed individually for three different assumptions for the primary mass composition: pure proton, pure iron and a simple two-component model. For each case, the unfolded spectra were obtained in three zenith angle bins. Assuming the isotropy of high energy cosmic rays reaching the Earth, one expects to obtain the same unfolded spectrum regardless of the zenith bin used for analysis. Although for pure proton and the two-component model the spectra obtained from three zenith angle bins were in good agreement, the pure iron case showed a strong disagreement between the three spectra  at low energies. It was concluded that pure iron primaries can be excluded below 24 PeV cosmic ray energies. This analysis puts the knee position around 4 PeV with spectral indices of 2.76 below and 3.11 above the knee. A flattening of the cosmic ray spectrum to an index of about 2.85 was also observed around 22 PeV.

The IceTop-73 analysis in figure~\ref{fig:specIC} is also IceTop-only analysis \citep{IT73}; no information from in-ice array is used. It is based on 11 months of data collected by 73 IceTop stations between June 2009 to May 2010 and covers the energy range 1-1000 PeV. This analysis is still in progress and systematic errors still need to be determined. However, as for the IceTop-26 analysis, the major source of systematic error is expected to be the unknown composition of cosmic rays; the analysis is done with proton and iron primary assumptions, individually. The final event sample has nearly 40 million events between 0.3 PeV and 1 EeV; about 200 events are above 200 PeV. The analysis is in agreement with the IceTop-26 analysis spectrum results.

The IceTop/IceCube-40 analysis  in figure~\ref{fig:specIC} comes from the first IceCube cosmic ray analysis that uses data from both surface (40 stations) and in-ice (40 strings) arrays \citep{IC40}. Only one month of data from August 2008 has been used in this analysis to demonstrate a method to measure the cosmic ray energy spectrum and composition at energies between 1 PeV and 30 PeV.  The analysis uses a neural network in conjunction with a $\chi^2$ minimization algorithm to get the average mass and energy of the cosmic rays. The input parameters are the S125 (a measure of air shower size) from the surface array and K70 (a measure of the muon bundle size) from the in-ice array. Within errors, the results of this analysis are in agreement with those from the IceTop-26 and IceTop-73 analyses. A power law fit yields the knee around 4.75 PeV with the spectral index being 2.61 below and 3.23 above the knee. 

\begin{figure}
\begin{center}
\includegraphics*[width=15cm,angle=0]{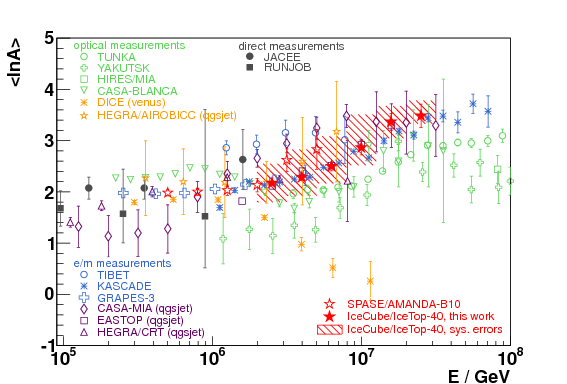}
\end{center}
\caption{IceTop/IceCube-40 analysis: Comparison of the mean logarithmic mass vs primary energy from different experiments; details are in the text.}
\label{fig:compKK}
\end{figure}

As shown in figure~\ref{fig:compKK}, the IceTop/IceCube-40 analysis also provides the average cosmic ray mass as a function of the reconstructed cosmic ray energy. The results are in agreement with SPASE-2/AMANDA-B10 \citep{SPASEAMA} and several other experiments that measure muon and electromagnetic components of the air showers. The analysis also shows a clear trend of increasing average mass with energy.

\begin{figure}
\begin{center}
\includegraphics*[width=15cm,angle=0]{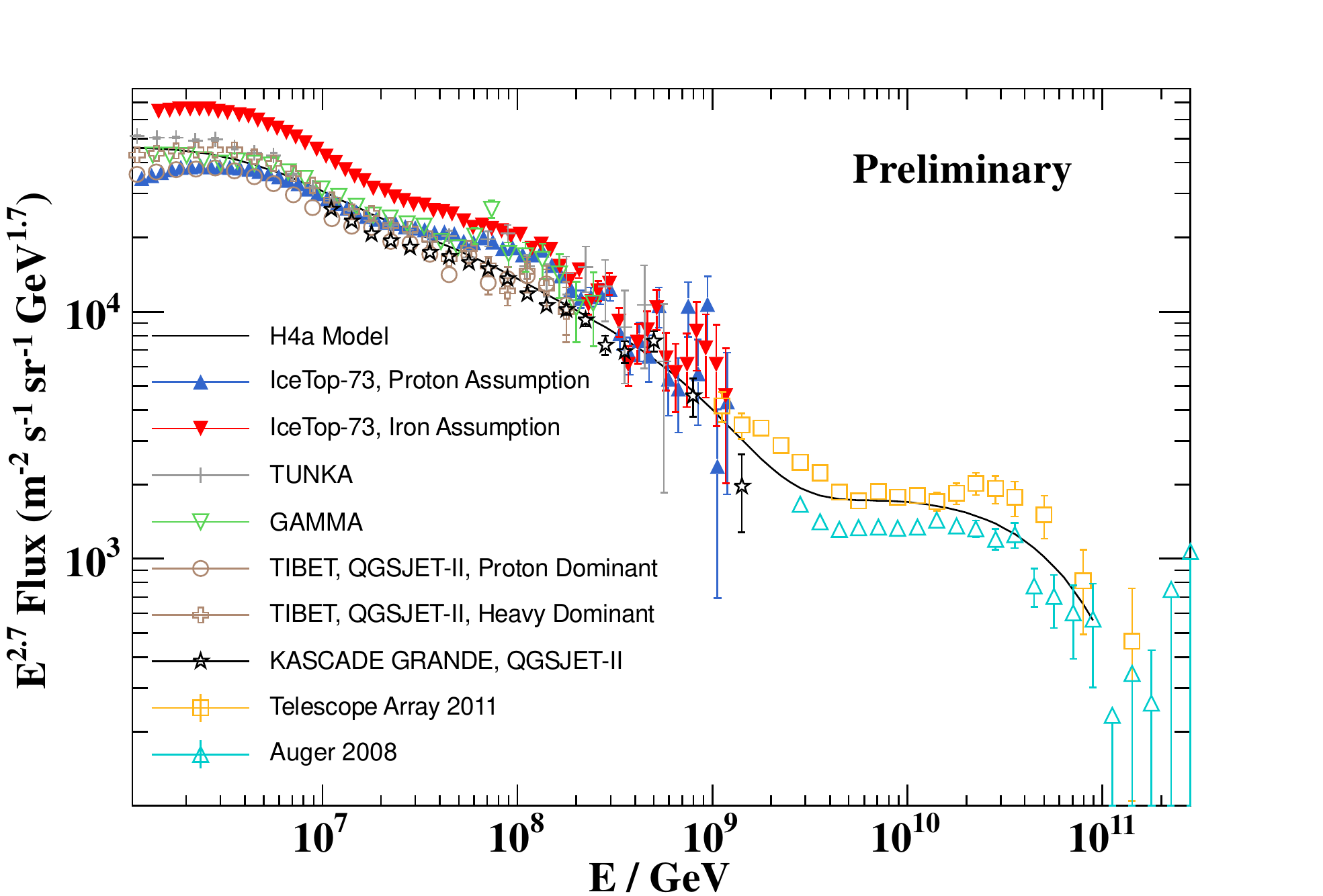}
\end{center}
\caption{Comparison of spectra resulting from IceTop-73 analysis with other experiments; black solid curve is the all particle spectrum from a three component cosmic ray model  \citep{H4a}.}
\label{fig:specAll}
\end{figure}

Figure~\ref{fig:specAll} compares the IceTop-73 all particle spectrum \citep{Alessio}, assuming pure proton (blue) and pure iron (red) primary cosmic rays, with the spectra obtained with recent measurements of KASCADE-Grande \citep{kascade}, Tibet Array \citep{tibet}, GAMMA \citep{gamma}, and Tunka \citep{tunka}. Also shown is a black solid curve labeled "H4a Model"; it is the all particle model spectrum \citep{H4a} that assumes three populations of cosmic rays (SNR component, high energy galactic component, and extra-galactic component). As we  can see in figure~\ref{fig:specAll}, a pure iron composition does not agree with the other experiments at energies below 20 PeV or so; this result is in agreement with the IceTop-26 analysis. Systematic error estimation is in progress for IceTop-73 analysis.

\section{Conclusions}

We have reviewed the IceTop/IceCube cosmic ray spectrum (between 1-1000 PeV) and mass measurement analyses performed so far. The results obtained are in good agreement among each other and with the other experiments. Further work is in progress to improve the systematics and composition sensitivity. Also, using the infill surface array that has a much smaller array spacing, we expect to go down to a few hundred TeV in cosmic ray energy; this will allow a comparison with direct measurements of cosmic rays. In future, a much larger statistics, improved understanding of the systematic uncertainties, and use of additional composition-sensitive parameters will be the major factors  that will contribute to a more precise measurement of the cosmic ray spectrum and its composition, especially in the knee region.

\clearpage

\end{document}